\documentclass[twocolumn,showpacs,amsmath,amssymb,prl,floatfix]{revtex4}

\usepackage{graphicx}
\usepackage{dcolumn}
\usepackage{bm}
\usepackage{bbm}
\usepackage{psfrag}
\usepackage{multirow,amssymb,amsbsy,amsmath}
\usepackage{stmaryrd}

\newcommand{\I}{\textup{i}}

\newcommand{\D}{\textup{d}}

\newcommand{\dd}{\text{d}}
\newcommand{\dod}[2]{\frac{\dd #1}{\dd #2}}

\newcommand{\pop}[2]{\frac{\partial #1}{\partial #2}}

\newcommand{\ddim}{\udelta\kern0.1em}

\newcommand{\beikonst}[2]{\left( #1 \right)_{\kern-0.2em #2}}
\newcommand{\tr}[2][]{\text{Tr}_{#1}\left\{#2\right\}}

\newcommand*{\bra}[1]{\mathopen{\langle}#1\mathclose{|}}
\newcommand*{\ket}[1]{\mathopen{|}#1\mathclose{\rangle}}

\newcommand{\comutxt}[2]{[#1,#2]}

\newcommand{\h}[1]{\hat{H}_{#1}}
\newcommand{\dop}{\hat{\rho}_A}

\newcommand*{\dbar}{\mkern 3mu\mathchar '26\mkern -12mu\D}
\newcommand{\dW}{\dbar W}
\newcommand{\dQ}{\dbar Q}
\newcommand{\heff}[1]{\hat{H}_{#1}^{\textup{eff}}}
\newcommand{\dheff}[1]{\dot{\hat{H}}_{#1}^{\textup{eff}}}
\newcommand{\eins}{\hat{1}}

\newcommand{\op}[1]{\hat{#1}}
\newcommand{\Exp}[1]{\text{e}^{#1}}
\newcommand{\e}{\Delta E}
\newcommand{\s}[1]{\hat{\sigma}^{#1}}
\renewcommand{\b}{\beta}

\begin{document}

%
%
\title{Local effective dynamics of quantum systems: A generalized
  approach to work and heat}


\author{Hendrik Weimer}%
\affiliation{Institute of Theoretical Physics I, University of Stuttgart, %
             Pfaffenwaldring 57, 70550 Stuttgart, Germany}%
\email{hweimer@itp1.uni-stuttgart.de}%
\author{Markus J. Henrich}%
\affiliation{Institute of Theoretical Physics I, University of Stuttgart, %
             Pfaffenwaldring 57, 70550 Stuttgart, Germany}%
\author{Florian Rempp}%
\affiliation{Institute of Theoretical Physics I, University of Stuttgart, %
             Pfaffenwaldring 57, 70550 Stuttgart, Germany}%
\author{Heiko Schr\"oder}%
\affiliation{Institute of Theoretical Physics I, University of Stuttgart, %
             Pfaffenwaldring 57, 70550 Stuttgart, Germany}%
\author{G\"unter Mahler}%
\affiliation{Institute of Theoretical Physics I, University of Stuttgart, %
             Pfaffenwaldring 57, 70550 Stuttgart, Germany}%




\date{\today}%

\begin{abstract}
  By computing the local energy expectation values with respect to
  some local measurement basis we show that for any quantum system
  there are two fundamentally different contributions: changes in
  energy that do not alter the local von Neumann entropy and changes
  that do. We identify the former as work and the latter as heat.
  Since our derivation makes no assumptions on the system Hamiltonian
  or its state, the result is valid even for states arbitrarily far
  from equilibrium.  Examples are discussed ranging from the classical
  limit to purely quantum mechanical scenarios, i.e. where the
  Hamiltonian and the density operator do not commute.
\end{abstract}


\pacs{05.30.-d, 03.65.-w, 05.70.Ln}

\maketitle

%
%


The formulation of classical thermodynamics was one of the most
important achievements of the 19$^{\textup{th}}$ century, as it
allowed to investigate a large variety of phenomena, including the
workings of thermodynamical machines. The first law of thermodynamics,
\begin{equation}
  \D U = \dW + \dQ,
\end{equation}
combined with definitions for the infinitesimal change in work $\dW$
and heat $\dQ$ and the second law is all that is required for
computing important quantities like the efficiency of a process.

In the quantum realm, the classification of work and heat is less
clear. So far, it has mainly been based on the change of the total
energy expectation value
\begin{equation}
  \label{eq:falsch}
  \D U = \D\tr{\h{}\hat{\rho}} = \tr{\hat{\rho} \D \h{} + \h{}\D\hat{\rho}},
\end{equation}
and defining the first term as $\dW$ and the second as $\dQ$
\cite{Alicki1979,Kosloff1984,Kieu2004,Henrich2006,Henrich2007}.
However, such a classification can be problematic, for various
reasons: For one, it is not obvious how to apply this definition to
processes involving an internal transfer of work and heat, as is the
case, e.g., in algorithmic cooling, a method to obtain highly
polarized spins by applying a series of quantum gates
\cite{Nielsen2000}. Then, the microscopic foundation of
(\ref{eq:falsch}) is rather unclear: As thermodynamic behavior may
occur even in small quantum systems \cite{Gemmer2004}, it should, in
principle, be possible to obtain $\dW$ and $\dQ$ even there. In the
following, we will present an alternative definition that does not
suffer from the problems above.

This letter is organized as follows. We first discuss the local
effective dynamics of a bipartite quantum system. This dynamics of one
part of the system is a reduced dynamics, which depend on the state of
and the interaction with the rest of the system.  Because we are
interested in local properties of a part of the system, it is
necessary to get a complete local description. In contrast to, e.g., a
Markovian quantum master equation (see, e.g., \cite{Breuer2002}) the
reduced dynamics cannot expected to be a set of closed differential
equations. Based upon what an experimentalist might observe, we give a
definition for the local energy. We then show that the change in this
local energy can always be split into a part that correlates with a
change in entropy and in a part which does not.  Corresponding to
classical thermodynamics, the former will be called ``heat'' and the
latter ``work''. However, this definition for the local heat and work
does not only depend on local properties, but on details of the whole
system. We explicitly give formulas to calculate those local
quantities once the time evolution of the full system is known.
Finally, two examples will be given, for illustration.

While work and heat can thus always be introduced to characterize the
energy exchange between two quantum subsystems, this splitting lacks,
in general, the robustness and universality, which is typical for the
corresponding thermodynamic concept. Additional investigations are
needed to show under which conditions this work and heat may be
identified with their thermodynamic counterparts. (This is reminiscent
of entropy, which also can be defined for any quantum state, but its
relationship to the thermodynamic variant requires further analysis.)
However, it is necessary to develop generalized
concepts for work and heat if one wants to analyze question requiring
both quantum mechanics and thermodynamics, like when investigating the
influence of coherence on the efficiency of a quantum machine.


We consider an autonomous bipartite system described by the
Hamiltonian
\begin{equation}
  \h{} = \h{A} + \h{B} + \h{AB},
\end{equation}
where $\h{A}$ acts on subsystem $A$ and $\h{B}$ on $B$, respectively.
We now focus on the local properties of subsystem $A$ only, however
our scheme would work also for subsystem $B$. The problem of local
addressability has been discussed intensively in quantum computing
\cite{Nielsen2000}. A local measurement of work or heat will have to
be be based on a coupling to a heat bath or work reservoir,
respectively. (The energy exchanged with a bath must be heat, that
with a ``mechanical'' reservoir work, by definition.) In both cases
the coupling will be realized by an interaction Hamiltonian $\h{AM}$
of the type
\begin{equation}
  \h{AM} = \sum\limits_i \hat{A}_i \otimes \hat{M}_i,
\end{equation}
where the operators $\hat{A}_i$ act only on subsystem $A$, and the
$\hat{M}_i$ only on the measuring device $M$. Choosing the operators
$\hat{A}_i$ may seem rather arbitrary, but can have important
consequences on the observed values. For example, let us consider a
system with an effective local Hamiltonian
\begin{equation}
\heff{A} = \frac{\Delta E}{2}\s{}_z.
\end{equation}
We now consider four cases: first, the system will be coupled to a
work reservoir, which can always be modeled by a classical driver, as
the work reservoir must always have zero entropy, by definition. We
use $\hat{A}_i = \hat{A} = g\sin(\omega t)\s{}_x$ and $\hat{A}=g\sin(\omega
t)\s{}_z$, respectively. For simplicity, we assume that the driving
frequency is off-resonant with respect to the eigenfrequency of
$\heff{A}$. Therefore, in the $\s{}_x$ case there will be no transfer
of energy, while in the $\s{}_z$ case $\D U$ is non-zero, and all
energy is transferred as work. We then repeat the process when
coupling the system to a heat bath. Here, we have $\hat{A} = \lambda
\s{}_x$ and $\hat{A} = \lambda \s{}_z$, respectively. In the $\s{}_x$
case, this results in system relaxing to a canonical state $\dop =
Z^{-1}\exp(-\beta \heff{A})$, with $\beta$ being the inverse
temperature. However, the $\s{}_z$ coupling results in a pure
dephasing of the system, i.e, $\D U = \dQ = 0$. A summary of the
possible combinations is shown in Table~\ref{tab:sxsz}.
\begin{table}
  \caption{Comparison of the scenarios when coupling a system to a
  heat bath or a work reservoir, depending on the interaction
  Hamiltonian.}
  \label{tab:sxsz}
    \begin{ruledtabular}
    \begin{tabular}{lll}
      & work reservoir & heat bath\\\hline
      $\s{}_x$ & $\D U = \dW = 0$ & $\D U = \dQ \ne 0$\\
      $\s{}_z$ & $\D U = \dW \ne 0$ & $\D U = \dQ = 0$\\
    \end{tabular}
    \end{ruledtabular}
\end{table}

The concrete physical realization thus defines a local effective
measurement basis (LEMBAS), which provides a reference for all
measurements of work or heat. Which choice of basis is the ``correct''
one, cannot be decided by the LEMBAS principle. However, there are
typically only a few choices which are physically reasonable, which
then can be further checked for consistency. A constructive procedure
that would allow to identify the correct basis is beyond the scope of
this letter.

In the following we do not consider the effect of the actual
measurement on the dynamics of the system. We only assume that some
fixed basis has been chosen as required by the LEMBAS principle, and
then compute work and heat with respect to this basis. Note that this
procedure is similar to a hypothetic von Neumann measurement -- the
probability for each outcome can be calculated without including the
measurement device in the dynamics.

We now define the infinitesimal work $\dW_A$ performed on $A$ as the
change in its internal energy $\D U_A$ that does not change its local
von Neumann entropy, i.e.
\begin{equation}
  \D S_A = 0 \Leftrightarrow \dW_A = \D U_A.
\end{equation}
The remainder is defined as the infinitesimal heat $\dQ$. It is
important to note that work and heat will then turn out to be
basis-dependent quantities as they depend on the choice of the
measurement basis.

The dynamics of the subsystem $A$ is given by the Liouville-von
Neumann equation
\begin{equation}
  \label{eq:vng}
  \pop{}{t} \dop = -\I\comutxt{\h{A}+\heff{A}}{\dop} + \mathcal{L}_{\textup{inc}}(\hat{\rho}),
\end{equation}
where $\dop$ is the reduced density operator of $A$, $\heff{A}$ is an
effective Hamiltonian (see below) describing the unitary dynamics
induced by $B$ on $A$, and $\mathcal{L}_{\textup{inc}}$ is a
superoperator describing incoherent processes, which may derive from
the environment of the total bipartite system (if present), but here,
in particular, from the influence of $B$ on $A$. Since
$\mathcal{L}_{\textup{inc}}$ is, in general, a function of the density
operator of the full system, Eq.~(\ref{eq:vng}) is typically not a
closed differential equation for $\dop$.

In the following we choose the energy basis of subsystem $A$ as the
measurement basis, so that only the parts of the total effective
Hamiltonian $\heff{A}$ that commute with $\h{A}$ contribute to the
described type of experiment. To find this part $\heff{1}$, we expand
$\heff{A}$ in the transition operator basis defined by the energy
eigenstates $\{\ket{j}\}$ of $\h{A}$:
\begin{equation}
  \heff{A} = \sum\limits_{jk} (\heff{A})_{jk} \ket{j}\bra{k}
\end{equation}
For this operator basis, we have
\begin{equation}
  \label{eq:proj-comms}
  [\ket{j}\bra{k},\h{A}]=\omega_{kj}\ket{j}\bra{k}
  ,
\end{equation}
where $\omega_{kj}$ is the difference between the energy eigenvalues
of the states $\ket{k}$ and $\ket{j}$, and therefore $\omega_{jj}=0$
for non-degenerate energy eigenvalues. Now, we define
\begin{equation}
  \heff{1} = \sum\limits_{j} (\heff{A})_{jj} \ket{j}\bra{j}
\end{equation}
which is the diagonal part of $\heff{A}$. From
Eq.~(\ref{eq:proj-comms}), we see that no non-zero linear combination
of transition operators with $j\neq k$ can ever commute with $\h{A}$
and therefore, neither $\heff{2}=\heff{A}-\heff{1}$, nor any part of
it commutes with $\h{A}$. Hence, $\heff{1}$ is the part contributing
to the measurement, as required, and we have
\begin{equation}
  [\heff{1},\h{A}] = 0,\;[\heff{2},\h{A}] \neq 0.
\end{equation}
The latter inequality holds except for the case where $\heff{2}=0$.
An analogous result can be achieved in the case of degenerate
eigenvalues of $\h{A}$ but has been omitted for the sake of brevity
and clarity.


If a measurement of the local energy is performed in the energy
eigenbasis of $\h{A}$ the corresponding operator is
\begin{equation}
  \h{A}' = \h{A} + \heff{1}.
\end{equation}
Therefore the change in internal energy within $A$ is given by
\begin{equation}
  \D U_A = \dod{}{t}\tr{\h{A}'\dop} \D t= \tr{\dot{\hat{H}}'\dop + \h{A}'\dot{\rho}_A}\D t.
\end{equation}
Using (\ref{eq:vng}) and assuming $\h{A}$ to be time-independent leads
to
\begin{equation}
  \label{eq:dUA}
  \D U_A = \tr{\dheff{1}\dop -\I\comutxt{\hat{H'}}{\heff{2}}\dop + \h{A}'\mathcal{L}_{inc}(\hat{\rho})}\D t,
\end{equation}
where the cyclicity of the trace has been used. Observing that the
dynamics induced by the first two terms is unitary, we arrive at
\begin{eqnarray}
  \dW_A &=& \tr{\dheff{1}\dop -\I\comutxt{\hat{H'}}{\heff{2}}\dop}\D t \label{eq:work}\\
  \dQ_A &=& \tr{\h{A}'\mathcal{L}_{inc}(\hat{\rho})}\D t.\label{eq:heat}
\end{eqnarray}
In this sense, it is possible to define heat and work for any quantum
mechanical process, regardless of the type of dynamics or the states
involved.


In order to actually compute $\dW$ and $\dQ$ the effective Hamiltonian
$\heff{A}$ is required. By starting with the Liouville-von Neumann
equation for the full system
\begin{equation}
  \pop{}{t} \hat{\rho} = -\I\comutxt{\h{}}{\hat{\rho}}
\end{equation}
and taking the partial trace over $B$ (see e.g. \cite{Breuer2002})
yields
\begin{equation}
  \pop{}{t} \dop = \tr[B]{\comutxt{\h{A} + \h{B} + \h{AB}}{\hat{\rho}}}.
\end{equation}
Using some theorems on partial traces shows that terms involving
$\h{B}$ vanish and $\h{A}$ generates the local dynamics in $A$. For
dealing with the terms involving $\h{AB}$ we first split the density
operator as
\begin{equation}
  \hat{\rho} = \dop\otimes\hat{\rho}_B + \hat{C}_{AB},
\end{equation}
where $\hat{\rho}_{A,B}$ are the reduced density operators for $A$ and
$B$, respectively, and $\hat{C}_{AB}$ is the operator describing the
correlations between both subsystems. Since the factorization
approximation is exact for the first term, we can write (cf.
\cite{Gemmer2001})
\begin{equation}
  \tr[B]{\comutxt{\h{AB}}{\dop\otimes\hat{\rho}_B}} = \comutxt{\heff{A}}{\dop},
\end{equation}
where $\heff{A}$ is given by
\begin{equation}
  \label{eq:heff}
  \heff{A} = \tr[B]{\h{AB}(\eins_A\otimes\hat{\rho}_B)}.
\end{equation}
Now we show that the processes generated by
$\comutxt{\h{AB}}{\hat{C}_{AB}}$ cannot result in unitary dynamics,
but will always change the local von Neumann entropy $S_A$. In order
to prove this, we convince ourselves that its time derivative is
non-zero, i.e,
\begin{equation}
  \dot{S}_A = -\tr{\comutxt{\h{AB}}{\hat{C}_{AB}}\log \dop\otimes\eins_B}\neq 0.
  \label{eq:entropy}
\end{equation}
Therefore, any dynamics generated by this term results in a
contribution to $\mathcal{L}_{inc}$. If the dynamics of the full
system was unitary, we would thus simply have
\begin{equation}
  \mathcal{L}_{inc} = -\I\tr[B]{\comutxt{\h{AB}}{\hat{C}_{AB}}}.
\end{equation}
as the total incoherent term.


How are these alternative definitions of heat and work linked to Eq.
(\ref{eq:falsch})?  It is easy to check that for $\heff{2} = 0$ the
definitions are compatible.  A pertinent example refers to a
quasistatic process within the system $A-B$ in thermal equilibrium at
temperature $T$. From the Gibbs fundamental relation it is known that
\begin{equation}
 \dd S_A=\frac{1}{T_A} \dbar Q_A.
 \label{eq:dS}
\end{equation}
Using now the definition given for the heat in (\ref{eq:heat}) we get
\begin{equation}
 \dd S_A=\frac{1}{T^*_A} \tr{ \h{A}^\prime \mathcal{L}_{\textup{inc}}(\hat{\rho}) } \dd t.
 \label{eq:dSheat}
\end{equation}
Here, $T^*_A$ specifies a parameter to be associated with the local
temperature. This parameter follows from the derivation of the entropy
$S_A$ from (\ref{eq:entropy}) combined with (\ref{eq:dSheat})
\begin{align}
 -\tr{\mathcal{L}_{\textup{inc}}(\hat{\rho}) \log \hat{\rho}_A } &= \frac{1}{T^*_A} \tr{\h{A}^\prime \mathcal{L}_{\textup{inc}}(\hat{\rho})} \notag \\
 T^*_A &= \frac{\tr{ \h{A}^\prime \mathcal{L}_{\textup{inc}}(\hat{\rho})}}{\tr{\mathcal{L}_{\textup{inc}}(\hat{\rho}) \log \hat{\rho}_A }}.
 \label{eq:temp}
\end{align}
For canonical states $\h{A}'$ commutes with $\dop$ and
$\mathcal{L}_{\textup{inc}}(\dop)$, and by using Eq.~(\ref{eq:dUA}) it
can be seen that Eq.~(\ref{eq:temp}) is equivalent to the classical
definition
\begin{equation}
  T^*_A = T_A = \pop{U_A}{S_A}.
\end{equation}
However, $T^*_A$ is not necessarily equal to the global temperature
$T$ of the full system due to the interaction between the individual
subsystems inducing correlations \cite{Hartmann2004,Hartmann2004a}.

Using the LEMBAS principle, it is now possible to investigate work and
heat also in non-standard physical scenarios. First we consider a two-level
atom with a local Hamiltonian $\h{A}$ interacting with a laser field
(subsystem $B$). In the semiclassical treatment of the radiation field
emitted by a laser the respective Hamiltonian is given by
\begin{equation}
  \label{eq:laser}
  \h{} = \h{A} + \heff{A} = \frac{\Delta E}{2}\hat{\sigma}_z + g\sin(\omega t)\hat{\sigma}_x,
\end{equation}
where $g$ is the coupling strength and $\omega$ the laser frequency.
In the rotating wave approximation the Hamiltonian can be made
time-independent. We investigate the situation where the atom is
initially in a thermal state described by the density operator
\begin{equation}
  \dop(0) = Z^{-1} \exp(-\beta\hat{H}_A),
\end{equation}
with $Z$ being the partition function and $\beta$ the inverse
temperature. The time-evolution of the density operator $\dop(t)$ can
be obtained by switching to the rotating frame and diagonalizing the
Hamiltonian.

Since (\ref{eq:laser}) is already an effective description for $A$, we
can directly compute $\dW_A$ and $\dQ_A$ once we know $\dop(t)$,
resulting in
\begin{eqnarray}
  \label{eq:dWrichtig}
  \dW_A(\delta,g) &=& \frac{\Delta E g^2}{2\Omega}\tanh \frac{\beta \Delta E}{2}\sin \Omega t\\
  \dQ_A &=& 0,
\end{eqnarray}
where $\Omega = \sqrt{g^2+\delta^2}$ is the Rabi frequency and $\delta
= \omega-\Delta E/\hbar$ is the detuning from the resonance
frequency. $\dW_A$ is the energy stored in $A$ that could be retrieved
after the preparing field has been switched off at time $\Omega
t=\pi/2$. For comparison, using the previously used definition for
the work [Eq.~(\ref{eq:falsch})] one was led to
\begin{equation}
  \dW_A(\delta,g) = \frac{(\Delta E+\delta) g^2}{2\Omega}\tanh \frac{\beta \Delta E}{2}\sin \Omega t.
\end{equation}
Comparison of the two results shows that in the latter case the
maximum is not at the resonance frequency ($\delta = 0$), i.e., only
our generalized approach is able to produce the correct physical
result.


\begin{figure}
\centering
 \includegraphics[width=8cm]{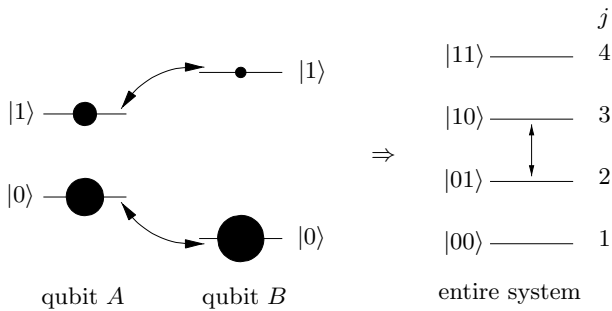}
 \caption{\label{fig:zweiSpinSWAP}Scheme of the SWAP gate and the accounting $\pi$-pulse.}
\end{figure}
As a second example we investigate the implications of our generalized
concept in the context of algorithmic cooling. Recently an algorithm
published by Boykin et al.~\cite{Boykin2002} has found experimental
realization by Baugh et al.~\cite{Baugh2005} and its thermodynamic
properties where investigated by Rempp et al.~\cite{Rempp2007}. Of
course, not all qubits [i.~e. two level systems (TLS) used for quantum
computing] are cooled down this way, some, the so called auxiliary
qubits, are heated up as well and thus have to be discarded or coupled
to the environment to work again as auxiliaries. The simplest system
to perform algorithmic cooling consists of two noninteracting TLS with
different Zeeman splitting ($\e_A < \e_B$) at the same initial inverse
temperature $\b_i$.  The cooling algorithm is a single SWAP gate,
which interchanges the occupation probabilities between the two spins
(see Fig.~\ref{fig:zweiSpinSWAP}); this implies a final inverse
temperature $\b_f=(\e_B/\e_A)\,\b_i$ of the cooled spin.  This
one-step cooling of a finite system (here one spin) reminds of a
refrigerator and thus should be characterized by properties like the
efficiency $\eta=Q/W$. To distinguish between work $W$ and heat $Q$
one needs to apply the local treatment described above. If one would
just use Eq.~(\ref{eq:falsch}) the heat would be $Q=0$ because the
entropy of the total system does not change. Therefore, it is
necessary to apply our generalized approach.

Again, in the semi-classical treatment of the subsystem $C$ inducing
the SWAP gate between the $\ket{01}$ and $\ket{10}$ levels the
Hamiltonian in the rotating wave approximation reads (index $j$ shown
in Fig.~\ref{fig:zweiSpinSWAP})
\begin{align}
 \op{H}(t)=&\op{H}_{AB}+\op{H}_{\text{int}}(t)
 =\sum_{j=1}^{4} E_j\,\op{P}_{jj}\nonumber\\
 +&\frac{1}{2}\,g\,\left(\Exp{\I\,(E_3-E_2)\,t}\op{P}_{23}+\Exp{-\I\,(E_3-E_2)\,t}\op{P}_{32}\right)
\end{align}
with $\sum_{j=1}^{4} E_j\,\op{P}_{jj}=\sum_{\mu =A,B} \e^{\vphantom{z}}_{\mu}\,\s{z}_{\mu}$ representing the spectrum of the two uncoupled spins.
To solve the equations of motion of this Hamiltonian we transform into the rotating basis to get rid of the time-dependence of the Hamiltonian \cite{Mahler1998}
\begin{equation}
 \op{H}_{\text{rot}}=\op{U}^{\vphantom{1}}_{\text{rot}}(t)\,\op{H}(t)\,\op{U}^{-1}_{\text{rot}}(t).
\end{equation}

Thus we are able to calculate the infinitesimal change of work
$\dW_{AB}$ done by the subsystem $C$ on the two-spin system by means
of Eq.~(\ref{eq:work}),
\begin{equation}
 \dW_{AB} =
\frac{g}{2}Z^{-1}\left(\Exp{\b_i\e_A}-\Exp{\b_i\e_B} \right)\big( \e_A-\e_B \big)
\sin{gt}
\end{equation}
where $Z$ is the partition function of the system.  To get the work
$W_{AB}$ induced by a SWAP we have to integrate over a half period of
the Rabi oscillation.  Because there is no detuning this is the same
result as if calculated directly via Eq.~(\ref{eq:falsch}). As in the
previous example $\dQ_{AB}=0$, i.e., subsystem $C$ only imparts
work. Interestingly enough, the situation changes when we only look at
$A$: To compute the transferred heat $Q_A$ we have to apply the LEMBAS
principle to investigate the local change of energy that is related to
a change of the local entropy.  We use (\ref{eq:heff}) to find the
effective Hamiltonian
\begin{align}
 \heff{A}=&-\s{0}\,\left\{\left[\Exp{\b_i\,(\e_A+\e_B)}-1\right]
\,Z^{-1}\e_B\right.\nonumber\\
&\left.+\left(\Exp{\b_i\,\e_B}-\Exp{\b_i\,\e_A}\right)
Z^{-1} \,\e_B\,\cos{(gt)}\right\}\nonumber\\
&+\s{z}\,\e_A.
\end{align}
From the fact that $\comutxt{\s{0}}{\s{z}}=0$ it follows that
$\heff{A}=\h{A}'$ and
\begin{equation}
 \dod{}{t}\op{\rho}_A(t)=\dod{}{t}\tr[B]{\op{\rho}(t)}=\mathcal{L}_{\text{inc}}.
\end{equation}
Using Eq.~(\ref{eq:heat}) yields
\begin{equation}
 \dQ_A=
-\frac{g}{2}Z^{-1}\left(\Exp{\b_i\,\e_A}-\Exp{\b_i\,\e_B} \right)\e_A\,\sin{(g\,t)},
\end{equation}
which, again, has to be integrated over a half period of the Rabi
oscillation to get $Q_A$. This is the heat that would have to flow
back from the original heat bath to let subsystem $A$ return to its
initial state. The SWAP gate thus has the engine efficiency $\eta_A$
\begin{equation}
 \eta_A=-\frac{Q_A}{W_{AB}}=\frac{\e_A}{\e_A-\e_B}.
\end{equation}
This is the same result as found for the quantum Otto process in the
heat pump mode \cite{Henrich2006}.

In summary we have shown the that exchange of energy between two
quantum systems can always be split into work and heat, based on local
effective dynamics. We have demonstrated how to obtain this local
effective dynamics, once the dynamics of the full system is known. As
this involves the choice of a measurement basis, those quantities are
basis-dependent. The choice of basis remains a subtle problem; a
detailed analysis of the environment should eventually give a unique
answer, just as in case of a measuring apparatus with respect to its
actual measurement basis. However, the examples presented show that
our generalized approach can lead to concrete and sensible results,
which is not always the case for the previously used one. The LEMBAS
principle should help to investigate thermodynamic aspects at the
borderline between full-fledged quantum dynamics and thermodynamics
proper.

We thank M.\ Michel, A. Kettler, G.\ Reuther, H.\ Schmidt, and J.\
Teifel for fruitful discussions. Financial support by the Deutsche
Forschungsgemeinschaft is gratefully acknowledged.


\end{document}